\documentclass[11pt,authoryear,3p]{elsarticle}
\usepackage{amssymb}
\usepackage{amsmath}
\usepackage{tabularray}
\UseTblrLibrary{booktabs}
\usepackage{multirow}
\usepackage{setspace}
\usepackage{xcolor}
\newcommand{\mbf}[1]{\mathbf{#1}}
\newcommand{\redd}[1]{\textcolor{red}{#1}}
\newcommand{\lgry}[1]{\textcolor{lightgray}{#1}}

\newcommand{\subnote}[1]{\begin{flushleft}\small\textit{Note:} #1\end{flushleft}}

\journal{arXiv}

\begin{document}

\begin{frontmatter}
\title{Conformity: Resolving the Trade-Off Between Performance and Synchrony in Multi-Unit Organizations}
\author{Ravshanbek Khodzhimatov\corref{cor1}}\ead{ravshan.personal@gmail.com}
\author{Stephan Leitner}\ead{stephan.leitner@aau.at}
\author{Friederike Wall}\ead{friederike.wall@aau.at}
\cortext[cor1]{Corresponding author}
\affiliation{organization={University of Klagenfurt},addressline={Universit{\"a}tsstra{\ss}e 65/67},postcode={9020},city={Klagenfurt},country={Austria}}

\begin{abstract}	
Multi-unit organizations are a form of organizations where the geographically dispersed units provide similar products or services in different markets. Deciding on an appropriate level of centralization in such organizations presents a unique challenge. One the one hand the organizations want to maintain a consistent brand identity in all units through centralized control, but on the other hand, they want to provide the units with sufficient autonomy to respond to the challenges they face locally. Traditionally, this challenge was perceived to require a trade-off between performance and organizational synchrony, with performance demanding more decentralization and synchrony requiring more centralized control. However, our research explores how organizations can potentially resolve this trade-off by promoting norms for knowledge-sharing and setting up the right communication channels, relying on the unit managers' intrinsic tendency to conform to the behavior of their peers. We build an agent-based model of an organization with multiple interdependent units facing highly similar task environments to investigate how unit managers' ability to communicate, share knowledge, and conform to peer practices might influence organizational dynamics. We find that, under specific communication network structures, increased decentralization can enhance both performance and organizational synchrony without sacrificing one or the other. Furthermore, we discover that centralization might still be preferable for synchrony if the units are interdependent.
\end{abstract}




\end{frontmatter}


\section{Introduction}
\label{sec:intro}

A multi-unit organization is a type of multidivisional organizations in which the units (divisions) provide a highly similar (if not the same) product or service in different locations, like retail chain stores or hotel branches. Having been rising for over a century \citep{kim99}, they have become a prominent structure in many industries, including retail, airlines, and banking \citep{greve2003,winter12}.

The multi-unit organizations constitute an interesting combination of a large corporation with many subsidiaries and a set of autonomous businesses operating in the same industry. Consequently, they face a unique challenge in determining the appropriate management strategy, specifically, the extent of (de)centralization in decision-making \citep{winter12}. On the one hand, organizations want all of their units to operate uniformly and to provide the same brand experience in all locations. For example, if one retail chain store discovers the optimal placement of shelves for their products, the upper management would naturally want to adopt this in the other stores \citep{jensen2007}. On the other hand, the upper management wants the units to perform as best as they can, adapting to the local environments, if necessary. For example, a retail chain store located in a suburban commercial zone can gain customers by offering free parking space, while this might be an unnecessary overhead for a store that is located in a walkable urban district \citep{eroglu1999}.

The existing studies consider the choice between unit performance and synchrony a necessary trade-off, where the former can be reached by giving more autonomy to the unit managers (i.e. decentralization) and the latter can only be enforced with strict guidelines and centralized control \citep{harrington00,garvin08}. However, most of these studies ignore the behavioral aspect of this problem \citep{gavetti2007} and assume that the unit managers are only interested in performance. In this paper we focus on an overlooked aspect of individual decision making, which is highly relevant to the question of synchrony --- \textit{conformity}. Conformity is the inner desire of individuals to match their own behavior to that of their peers \citep{cialdini04}. Studies in the field of psychology state that individuals conform for various reasons, including emotional ones like fitting into a group or being consistent with their perceived self-image, and more strategic ones like taking the behavior of others as a reference point to increase their own accuracy in cases when the decision they are facing is too difficult or not obvious to solve on their own. Conformity is generally understood to be an inner motivation that cannot be promoted externally. However, the organization can influence this process by promoting the culture of sharing knowledge \citep{quigley2007,argote2020} and by setting up communication channels between the unit managers \citep{tsai2001,lee2019}.

We conjecture that if the unit managers conform to their peers, this can result in an increase in organizational synchrony even without the centralized directive, while keeping the benefits of the decentralization to the organizational performance, thus resolving the trade-off. To check this idea we build an agent-based model of a multi-unit organization in which several agents (unit managers) work on highly similar (sometimes interacting) sets of tasks, communicate with each other through different network structures, and conform to the behavior of peers they communicated with. We simulate organizations with different levels of centralization, and check if the decentralized decision-making with conformity can lead to an increase in both performance and synchrony.

\section{Theoretical Motivation}
\label{sec:litreview}

In this section we present theoretical motivation for our research. We start with describing multi-unit organizations and the trade-off between performance and synchrony that they face. We then discuss how choosing the appropriate level of (de)centralization has been used to address this trade-off. Finally, we introduce the concept of conformity and how it can be used along with norms for knowledge sharing and the appropriately designed communication channels to potentially improve both performance and synchrony, thus resolving the trade-off.

\subsection{On the structure of multi-unit organizations}
\label{sec:litmuo}

Multi-unit organizations are defined as dispersed organizations built from standard units such as branches and stores, each offering the same product or service \citep{garvin08}. They can be considered a subclass of multidivisional (or M-form) organizations \citep{buck1985}, with the main difference in that they do not have divisions providing complementary or unrelated products or services \citep{greve2003}. It is important to notice, though, that the units are not independent businesses, they are still part of a single organization and have a common overarching strategy and communication channels with other units \citep{maness1996}. Thus, in practical terms, multi-unit organizations can also be thought of as the collection of several franchise stores with the condition that they belong to a single franchisee \citep{darr1995,michael2000}.

One of the primary advantages of the multi-unit organizations is that they allow for faster growth through direct replication to new locations of a template that worked well in the existing units \citep{winter12,audia2001,darr1995}. Another advantage is that having branches in different locations allows organizations to engage in strategic behavior not available to organizations with units operating in different markets, like competing at a global scale or influencing prices in the entire country \citep{greve2003}. Furthermore, multi-unit organizations can benefit from the \textit{economies of marketing} by using a single advertising campaign for all branches \citep{kim99} or introducing loyalty programs that can be redeemed in many locations \citep{greve2003}.

In general, with a few exceptions \citep{harrington00}, multi-unit organizations are highly centralized, featuring multiple tiers of management -- local (or unit-level), regional, and headquarters, where the higher level management usually determines the strategy and the lower level implements it, although the responsibilities can significantly overlap \citep{garvin08}. There are several reasons that can explain this tendency to have a centralized control. One reason is the concern that autonomously operating units might fail to coordinate and potentially miss out some important interactions between units \citep{ethiraj04b} or even start to compete with each other \citep{seran2023}. Another, more important reason is the willingness to replicate successful templates as precisely as possible because the imperfect replication might be detrimental to the organization due to the failure to capture the interlocking practices \citep{jensen2007,winter12} and to operate in \textit{synchrony}.

\subsection{Synchrony as a goal of multi-unit organizations}
\label{sec:litsync}
By \textit{synchrony} in this paper we refer to the overall similarity of products or services provided by the organizational units at any given point in time, echoing the concept of \textit{behavioral synchrony} in psychology, which denotes the act of keeping together in time with others \citep{baimel15}. By design, multi-unit organizations always aim to achieve synchrony, because they promise customers the same brand experience and a uniform image in all of their units \citep{garvin08}.

One of the main reasons as to why organizations might want to offer similar service in all of their units is the desire to ensure that the best practices discovered by any branch would be immediately adopted by the others to \textit{exploit} the success of these practices \citep{march91,darr1995,baum1998}. Indeed, the organization's ability to effectively transfer knowledge from one unit and make it adopted by another is seen as an important source of competitive advantage \citep{argote2000,miller06,haldin00} and a main form of growth for organizations such as Walmart or Starbucks \citep{winter12}. Another reason for organizations to pursue synchrony includes making sure that the unit managers can fully benefit from the organization's standardized guidance and support, which may not be possible if the units deviate significantly from the established template \citep{winter12}. Furthermore, the best practices that the units are expected to follow might be carefully interlocked with each other, and modifying complex but imperfectly understood templates might negate their benefits \citep{jensen2007}. Finally, it was found that the large franchise organizations that failed to enforce the uniformity in their units faced an increased risk of unit failure \citep{winter12}.

It naturally follows from this discussion that before replicating a succesful practice or template, one must first develop it \citep{winter01,argote2020,ding2010}. Moreover, the different locations in which the units operate might feature different market conditions or customer preferences, and require some level of customization or local adaptation \citep{garvin08,winter12}. This means that the organizations have to concede some level of autonomy to the unit managers (i.e. decentralize) to allow them to react to their local environments.

\subsection{Balancing synchrony and performance with (de)centralization}\label{sec:tradeoff}

\citet{argote2020} highlighted organizational search (for the best practices or solutions to the problems that the organization faces) as the first fundamental process of organizatinal learning (see also \citet{cyert1963} and \citet{nelson1982}). It is closesly related to the concepts of \textit{exploration} and \textit{exploitation}, where the former refers to the broad search for the novel practices and the latter refers to the refinement and slow improvement of the existing ones \citep{march91}. The advantages of synchrony mentioned in the previous section mostly cover the exploitaiton, but in order to successfully operate in different markets and conditions the multi-unit organizations are also interested in the exploratory search \citep{sorenson2001}.

Exploration of novel practices poses a substantial challenge and cannot be simply commanded by the upper management. On the contrary, the chances to discover significantly better performing practices are increased when the organizations decentralize their search efforts \citep{sorenson2001,ethiraj04b,kocak2023}. This can be explained by the broader coverage of \textit{parallel search}, in which multiple units search simultaneously for better solutions to the same problem \citep{baumann18}. Another explanation is that in the absence of a constant need to validate the decisions with the upper management or the peers, the units might experiment more frequently until they discover better performing solutions \citep{csaszar2013,harrington00}.

Similary, achieving synchrony, once the high performing practices had been discovered, is not a trivial task and requires following the knowledge creation and transfer process \citep{argote2000,argote2020}. This mainly includes the \textit{codification} of findings of a unit into routines or principles by the upper management and the spread of these routines to other units \citep{argote2000}. In general, these processes run more efficiently in centralized organizations with vertical communication to all units \citep{alonso2008,giannoccaro2004,harrington00}. However, in some cases the knowledge can be \textit{tacit}, i.e. non-codifiable into guidelines or routines, and can only be spread via direct decentralized communication between the units \citep{haldin00,miller06,hansen2002}. 

To sum up, organizations must constantly balance between the search for the better solutions (goal of performance) that requires some level of decentralization, and the adoption of these solutions in all of the branches (goal of synchrony) that requires some level of centralized control. However, this trade-off between performance and synchrony must not be considered as a strict binary choice between centralization and decentralization. Firstly, (de)centralization lies on a spectrum which includes not only purely vertical hierarchies and fully autonomous units, but also the units with voting or vetoing power \citep{csaszar2013}, the power to pre-screen alternatives before upper management \citep{siggelkow05a}, or temporary decentralization followed by reintegration \citep{siggelkow2003}. Secondly, the reasoning that knowledge transfer is not possible without a centralized intervention, implicitly assumes that the individuals (unit managers) are only interested in the unit performance, while a substantial body of research in psychology suggests that the individuals themselves want to \textit{conform} to their peers and adopt their behavior \citep{cialdini04}.

\subsection{Knowledge sharing and conformity in organizations}
\begin{figure}
	\centering
	\includegraphics[width=0.75\linewidth]{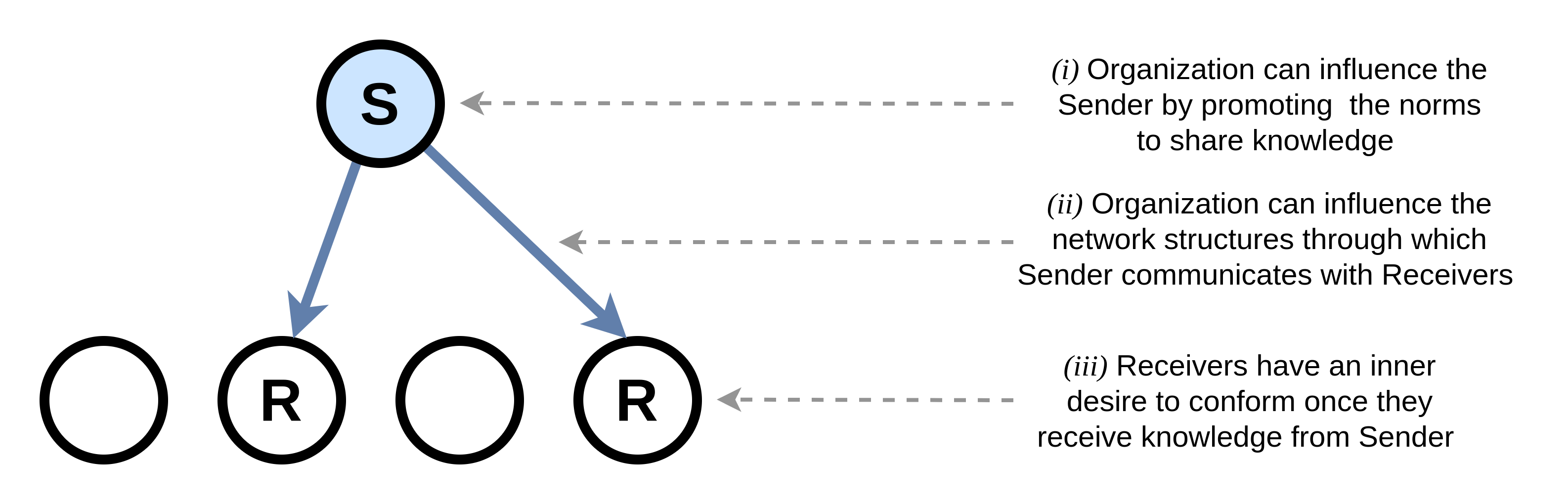}
	\caption{Conformity and knowledge sharing}
	\label{fig:orgconformity}
	\subnote{Conformity and knowledge sharing consists of Sender unit managers (S), Receiver unit managers (R), and the communication channels between the units.}
\end{figure}

There have been numerous studies in the field of organizations that focused on the formal mechanisms (e.g. coordination modes, incentive schemes, organizational structures) to achieve organizational goals, and significantly fewer studies that incorporated insights from psychology regarding indvidiual behavior \citep{gavetti2007}. We will attempt to address this gap and see how the individuals' inner willingness to \textit{conform} \citep{cialdini04} might affect the transfer of knowledge and, as a consequence, synchrony in the organization. We decompose this process into three main aspects --- sharing of knowledge by unit managers, communication between unit managers, and the adoption of the received knowledge, as illustrated in the Figure \ref{fig:orgconformity}. Next, we will discuss these three aspects in more detail and determine how they can be influenced to achieve synchrony.

\paragraph{(i)} The willingness to share knowledge sets the stage for the entire knowledge transfer process by defining the content of knowledge that is going to be shared, communicated and adopted \citep{argote2020}. It starts with the individuals (unit managers) explicitly sharing their practices with their peers or deliberately making them observable \citep{chan2014}. This behavior is not necessarily inherent to all individuals, and might require some prompting by the organization \citep{seran2023} such as promoting norms for sharing knowledge \citep{quigley2007}, instituting a culture of cooperation and raising reputational costs of not sharing \citep{argote2020}, introducing fixed or team-based incentive schemes \citep{lee2017}, considering power and status differences in the formation of teams \citep{bunderson2011,tsai2002}. However, there are some factors that compel individuals to share knowledge without external intervention, such as positively identifying themselves with their peers belonging to the same demographic group \citep{hansen2002} or sharing the same culture \citep{distefano2014}, not being able to resist telling about their good ideas to their colleagues, even including the competitors \citep{baum1998}.

\paragraph{(ii)} Once the individuals (unit managers) decide to share the knowledge, it goes to their peers with whom they are connected via various communication channels. While some individuals might be connected informally (e.g. through personal acquaintance), it is, in general, a task of organization to design and promote the communication channels that would lead to the efficient transfer of knowledge. There are many mechanisms that organizations can use to achieve this, such as organizing training programs \citep{argote2020} and informal meetups \citep{tsai2002,darr1995,orlikowski2002}, setting up instant messaging applications \citep{guilbeault2018}, facilitating inter-firm mobility \citep{marino2016}, and even the closer arrangement of seats in the office if managers work there \citep{lee2019,catalini2018}. These communication channels, usually referred to as the \textit{network structures}, have been subject to extensive studies, with some relevant insights including that organizational units centrally positioned in the network can have access to more innovative knowledge \citep{tsai2001}, organizations with too many \textit{hubs} (centrally positioned units) do not perform well, efficient networks lead to high synchrony but the inefficient networks lead to a diversity in discovered solutions \citep{lazer2007}, and networks with semi-isolated sub-groups lead to the best balance of high performing solutions and the synchrony in the organization \citep{fang2010}, and, finally, organizational networks remain relatively stable over time unless intervened upon \citep{kilduff2003}.

\paragraph{(iii)} Finally, once the Sender individuals share their knowledge and spread it through the network, the Receiver individuals gain access to that knowledge and make a decision whether to adopt it or not. This process is subject to limited control by an organization and is governed more by the individuals' inner willingness to conform. Conformity can be defined as the act of changing individual's own behavior to match that of others \citep{cialdini04}. Research in the field of psychology generally identifies two main motives for conformity -- informational and normative, where the former refers to the desire to find the most accurate information in the behaviors of others, while the latter refers to the willingness to fit into a group or gain social approval \citep{deutsch55}. These two motives can also be categorized as the goal of accuracy and goal of affiliation respectively \citep{cialdini04}. There are situations in which the goal of affiliation takes precedence and compels the individuals to conform to the peers they positively identify with, even if it decreases their performance \citep{cialdini90,kane2005,gali1994,charroin2022}, however in the context of multi-unit organizations, in which all units are providing the same product or service and are trying to perform well, we will mostly consider that they conform to achieve the goal of accuracy. The main reasoning here is seeing others' decisions as a reference point \citep{tversky1991} when the environment features a \textit{causal ambiguity} and the individuals cannot observe the outcomes of their decisions immediately and clearly, thus making behaving exactly as the others to be perceived as reducing the risks of the wrong decision \citep{winter12,gavetti2007,buckert2017}. Other reasons include the low self-efficacy (i.e. trust in one's own abilities) \citep{quigley2007} and the high cost of mistakes \citep{bandura71,doucouliagos1996}. There are several factors that limit the conformity, such as low \textit{absorptive capacity}, e.g., when a unit manager does not have enough infrastructure to immediately adopt all practices of another unit \citep{tsai2002}, or the \textit{complex contagions} aspect, in which individuals do not immediately conform to others, and wait until they encounter the same practices in several peers \citep{guilbeault2018}.

In this paper we will focus on the question of whether this decentralized way to achieve synchrony can be effective, its effect on the organizational performance, and how it compares to the more traditional centralized way in achieving the balance of performance and synchrony in multi-unit organizations. For the purpose of this study, we omit the costs of promoting the norms for sharing, the costs of establishing the communication channels, and the limitations to the individual conformity. While these factors are undoubtedly important, they would introduce additional complexity to our model, and leaving them to future research would allow us to maintain a stronger focus on our main research question.



\section{Model}
\label{sec:model}

We approach the research question using an agent-based model with two types of agents (unit managers and upper management). This allows us to simulate communication and knowledge sharing between the unit managers and to compare different levels of centralization in the multi-unit organizations. In this section we summarize the main elements of the model, including the task environment and the network structures, and define the agents and their motivations. We include an in-depth description of the simulation design and the detailed calculations and expansions of the presented formulas in the Online Appendix 1 following the \textit{Overview, Design concepts and Details} (ODD) protocol \citep{grimm2020}.

\subsection{Task environment}
\label{sec:taskenv}

We model the organization using the $N\!K$-framework \citep{kauffman89,levinthal97}, which uses the concept of \textit{performance landscapes} to represent the environment in which the organizations operate. The landscape constitutes a mapping from a set of interdependent (or \textit{coupled}) decisions to a numeric value that denotes the performance. The decisions are modeled as binary (i.e. taking values of $0$ or $1$) and the performances are drawn from uniform random distribution. This allows us to focus on how agents search for better solutions for a given generic problem without specifying the details. What we do specify is the number $N$ of tasks each agent faces and the number $K$ of tasks that are coupled with (i.e. affect the performance of) each task. Hence, the organization with $P$ units, each consisting of $N$ binary tasks can be represented as:
\begin{equation}
	\left( \underbrace{x^1_1, ..., x^1_N}_{\mathbf{x}^1}, \underbrace{x^2_1, ..., x^2_N}_{\mathbf{x}^2}, ..., \underbrace{x^P_1, ..., x^P_N}_{\mathbf{x}^P} \right)
\end{equation}

In our model of the multi-unit organizations we want to capture that the units might be (though, not necessarily) interdependent and have coupled tasks. To achieve that we use the $N\!K\!C\!S$-model \citep{kauffman91}, which is a variation of the $N\!K$-model where each agent (unit manager) faces a performance landscape with each task coupled with $K$ other tasks from the same landscape and $C$ tasks from each of the $S$ other agents' landscapes, a total of $K + C\cdot S$ couplings for each task. Thus, we can denote the performance of a binary decision $x^p_i \in \{0, 1\}$ on task $i$ of agent $p$, drawn from a uniform distribution, by:
\begin{equation}
	\phi (x^p_i, \underbrace{\dots}_{K}, \underbrace{\dots}_{C \cdot S}) \sim U(0,1)
\end{equation}
Moreover, we want to capture that each unit is facing a highly similar but not necessarily identical problem. To do that we generate correlated landscapes with an equal number of tasks, $N$, and assign them to the agents \citep{verel13}. Implementation wise, this means that the performances of all the corresponding tasks $i \in \{1,...,N\}$ of different agents $p$ and $q$ are correlated with a constant coefficient $\rho$.	

Having defined the performance of a task, we straightforwardly calculate the unit performance $\phi^p$ as the mean performance of the tasks assigned to the unit $p$:
\begin{equation}
	\phi^p = \frac{1}{N} \sum^N_{i=1} \phi(x^p_i),
\end{equation}
and the organizational performance $\Phi$ as the mean of unit performances:
\begin{equation}
	\Phi = \frac{1}{P} \sum^P_{p=1} \phi^p
\end{equation}

Finally, we want to check the extent of similarity between the decisions on corresponding tasks of different units at every point in time. To do that, we define a metric for \textit{synchrony} that compares the decisions on corresponding tasks between all pairs of agents, calculates how frequent these decisions are distinct, and takes a complement of this value.
\begin{equation}
	\mathbf{S}  = 1 - \frac{1}{\mathbf{H}^{\max}_{P,N}} \sum_{p=1}^{P} \sum_{q=p+1}^{P} H(p,q),
\end{equation}
where $H(p,q) \in \{0,..., N \}$ is the number (count) of distinct decisions on corresponding tasks of agents $p$ and $q$, $\mathbf{H}^{\max}_{P,N}$ is the maximum possible number of pairwise distinct values for given $P$ and $N$, and the value of $\mathbf{S}$ falls between $0$ and $1$.\footnote{The calculation of this formula is given in more detail in Online Appendix 1.}

\subsection{Communication and conformity}
\label{sec:networks}

\begin{figure}
	\includegraphics[width=\linewidth]{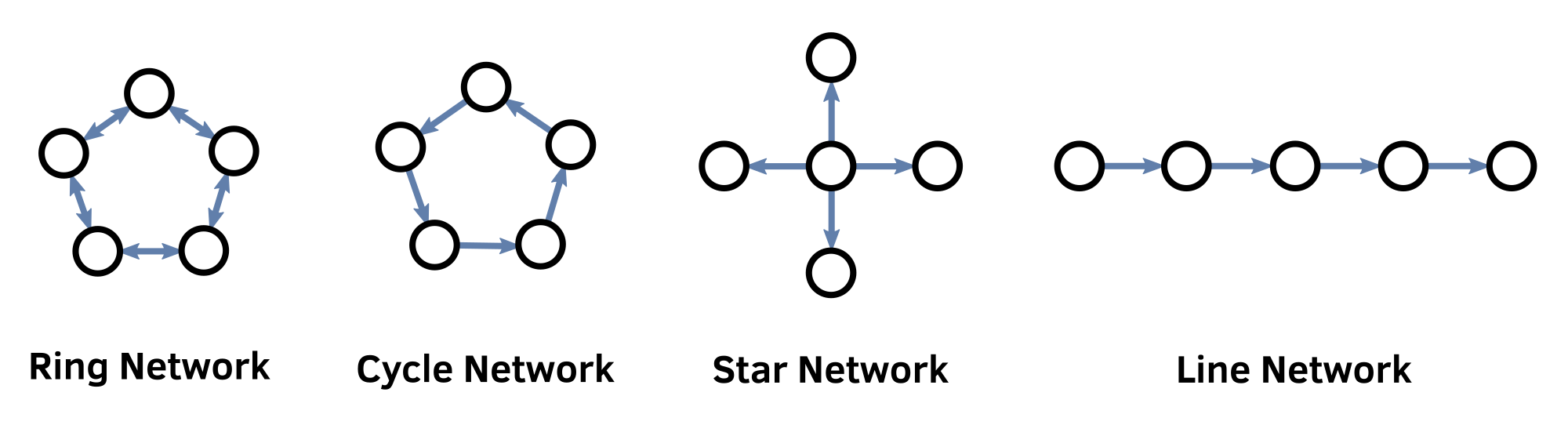}
	\caption{Network structures}
	\label{fig:network}
	\subnote{Network structures in which agents (nodes) share information through directed links. Ring and Cycle networks feature the decentralized flow of information, while Star and Line networks feature the top-to-bottom centralized flow of information. Moreover, the Ring and Star networks represent the fast transfer of information due to the direct connectedness to multiple units, while the Cycle and Line networks represent the slow transfer of information in which each agent is connected to only one other agent.}
\end{figure}

The agents (unit managers) communicate and share their most recent decisions $\mathbf{x}^p = (x^p_1, ..., x^p_N)$ with each other via one of the 4 stylized network structures illustrated in Figure \ref{fig:network}. The Ring and Cycle network structures feature a decentralized flow of information, while the Star and Line network structures feature a centralized flow of information starting from a single node (unit manager).\footnote{Note that the (de)centralized flow of information is categorically different from the (de)centralized organizational structure discussed in this paper.} At the same time the Star and Ring networks feature more direct connections between nodes, which leads to a faster flow of information, while the Cycle and Line networks require the information to pass through other nodes before reaching the receiver, which leads to a slower flow of information.\footnote{For a more detailed analysis of these four network structures consult \citet{khodzhimatov2023}}

Once the agents receive the information about each other's decisions, they store it in their memory for $T_M$ periods after which they \textit{forget} it.\footnote{See \citet{argote2020} for knowledge depletion and forgetting, and \citet{xie02} for the algorithm that features forgetting} The agents use the information in their memory $M$ to calculate the extent of conformity $\kappa$ of the set of decisions $\mathbf{x}^p = (x^p_1, ..., x^p_N)$ as:
\begin{equation}
	\kappa (\mathbf{x}^p) = \frac{1}{|M|} \sum_{\mathbf{x} \in M} I(\mathbf{x}^p, \mathbf{x}),
\end{equation}
where $I(\cdot, \cdot)$ is the fraction of similar decisions on the corresponding tasks of the two arguments, $|M|$ is the number of entries in the memory, with each entry $\mathbf{x}$ consisting of $N$ decisions.

\subsection{Agents' objectives}
\label{sec:agents}

Having defined the performance, synchrony, and conformity, we are ready to present the objective functions of the agents. 
The unit managers want to increase their utility defined as the weighted sum of performance and conformity when making their decisions on the $N$ tasks assigned to them:
\begin{equation}
	\max \ \  w_\phi \cdot \phi^p + w_\kappa \cdot \kappa,
	\label{eq:utility}
\end{equation}
where $w_\phi$ and $w_\kappa$ are the weights for unit performance and measure of conformity respectively. The upper management wants to attain the goal of performance, $g_\Phi$, and the goal of synchrony, $g_\mathbf{S}$, by minimizing the following function:
\begin{equation}
	\min \ \  d^{-}(\Phi, g_\Phi) + d^{-}(\mathbf{S}, g_\mathbf{S}),
	\label{eq:goals}
\end{equation}
where $d^{-}(\cdot,\cdot)$ denotes the underachievement of the goal, and is equal to the difference of the goal and the value if the goal is not achieved, or is equal to zero if the goal is achieved or exceeded.

\subsection{Decision-making and coordination}
\label{sec:coordination}

The task environment the agents are facing is complex by design and requires a search procedure to find the best solutions. As a baseline approach we take the \textit{hill-climbing} (or \textit{local adaptation}) algorithm \citep{levinthal97} in which the agents start by setting a random set of decisions as their status quo and alter one decision at a time. If this increases their utility, they \textit{climb up} by declaring this altered set of decisions as their new status quo. Otherwise, they stick to the current set of decisions. This algorithm is commonly used in studies that utilize the $N\!K$ framework \citep{wall16} with the caveat that it does not always reach the highest possible value (global optimum) of the objective function (getting stuck in the local optima). Moreover, when the tasks of different agents are interdependent, making unilateral decisions is suboptimal, as it fails to capture the collaborative solutions that might be better for all agents.

In this paper we consider three coordination modes built on top of the hill-climbing algorithm --- Decentralized, Pre-screening, and Centralized, as introduced by \citet{siggelkow05a} and summarized in Table \ref{tab:coordmode}. All coordination modes begin with agents (unit managers) looking at $ALT$ number of alternative sets of decisions \textit{in the neighborhood of} (i.e. differing in only one decision from) the status quo.

\paragraph{Decentralized.} In the decentralized coordination mode, the agents look at $ALT=2$ (or $ALT=4$ if they search thoroughly) alternatives and climb to the one with the highest utility (see Equation \ref{eq:utility}) if it is greater than the current value, or stick to the status quo otherwise. The upper management does not participate in decision-making in this coordination mode.

\paragraph{Pre-screening.} In the pre-screening coordination mode, the agents look at $ALT=4$ alternatives and pick the $PROP=2$ ones with the highest utility (see Equation \ref{eq:utility}) and submit them to the upper management, which then randomly picks $COMP=2$ (or $COMP=4$ if it searches thoroughly) combinations of the submitted proposals from all agents and makes a climbing decision based on the attainment of goals (see Equation \ref{eq:goals}).

\paragraph{Centralized.} The centralized coordination mode is similar to the pre-screening mode, except the agents randomly pick $ALT=2$ decisions and submit them to the upper management without pre-screening, making the entire climbing decision depend on the attainment of organizational goals and not the individual utilities of agents. In line with the literature on multi-unit organizations we consider this a default coordination mode \citep{garvin08}.

\begin{table}[!tb]
	\small
	\centering
	\caption{Coordination modes}
	\label{tab:coordmode}
	{
		\footnotesize
		\begin{tblr}{colspec={lXX}, row{1}={c,ht=1.5em}}
			\toprule
			& \bf Cursory search & \bf Thorough search \\
			\midrule
			
			\bf Decentralized & Unit managers screen $ALT=2$ alternatives and decide to switch to the highest or not. & Unit managers screen $ALT=4$ alternatives and decide to switch to the highest or not.  \\ 
			
			\bf Pre-screening & Unit managers screen $ALT=4$ alternatives, pick $PROP=2$ best (including the current configuration), and submit to the upper management, who randomly combines the proposals into $COMP=2$ combinations and decides to switch to one of them or not. & Unit managers screen $ALT=4$ alternatives, pick $PROP=2$ best (including the current configuration), and submit to the upper management, who randomly combines the proposals into $COMP=4$ combinations and decides to switch to one of them or not.   \\
			
			\bf Centralized & Unit managers randomly pick $ALT=2$ alternatives and submit all of them ($PROP=2$) the upper management, who then randomly combines the proposals into $COMP=2$ combinations and decides to switch to one of them or not. & Unit managers randomly pick $ALT=2$ alternatives and submit all of them ($PROP=2$) to the upper management, who then randomly comnbines the proposals into $COMP=4$ combinations and decides to switch to one of them or not. We consider this a default coordination mode in multi-unit organizations. \\
			
			\bottomrule
		\end{tblr}
	}
\end{table}

\subsection{Simulation design}
\label{sec:process}

To sum up, we model a multi-unit organization as agents operating on interdependent performance landscapes, where the unit managers communicate with each other via the network structures and share information about their decisions, and make choices to increase their utility, while the upper management wants to attain the goals of performance and synchrony. Figure \ref{fig:process} in the Appendix \ref{sec:appendixA} presents a flow chart diagram of the sequence of events happening at each time step of our simulation.

Before running the simulations, we construct scenarios which include 2 task environments with and without interdependencies between landscapes, 4 network structures through which the agents communicate, 3 coordination modes and 2 search modes for decision making. Table \ref{tab:params} specifies the parameters used to construct these scenarios.\footnote{The choice of parameters for task environment has been informed by \citet{khodzhimatov2024} and the choice of parameters for coordination and search modes has been informed by \citet{siggelkow05a}.}

\begin{table}[!tb]
	\small
	\singlespacing
	\caption{Simulation scenarios}
	\label{tab:params}
	\begin{tblr}{X[-1,c]X[l]X[-1,l]}
		\toprule
		\bf Parameter & \bf Description & \bf Value \\
		\midrule
		$P$       & Number of agents                            & $5$  \\
		$N$       & Number of tasks assigned to a single agent  & $4$  \\
		$(K,C,S)$ & Interdependence structure			        & $(3,0,0)$, $(2,2,2)$ \\
		$\rho$    & Pairwise correlation between landscapes     & $0.9$ \\
		$T_M$     & Memory span of agents                       & $50$ \\
		$(w_\phi, w_\kappa)$  & Weights of performance and conformity       & $(0.5, 0.5)$ \\
		$(g_\Phi,g_\mathbf{S})$ & Goals of performance and synchrony        & $(1.0, 1.0)$ \\
		  & Network structure    						& Ring, Cycle, Star, Line \\
		  & Coordination mode    						& Decentralized, Pre-screening, Centralized \\
		  & Search mode    								& Cursory, Thorough \\
		$ALT$     & Number of screened alternatives             & $2$, $4$ \\
		$PROP$    & Number of submitted proposals               & $2$ \\
		$COMP$    & Number of proposal combinations             & $2$, $4$ \\
		$T$       & Observation period                          & $500$ \\
		$R$       & Number of simulation runs per scenario      & $1000$ \\
		\bottomrule
	\end{tblr}
	\subnote{The coordination modes determine what the values for ($ALT$, $PROP$, $COMP$) stand for. For example, a thorough search is defined by $ALT=4$ for decentralized mode and by $COMP=4$ for pre-screening or centralized mode.}
\end{table}

For each scenario we run $R=1000$ simulations and in each of them we observe the normalized organizational performance\footnote{The performance landscapes are generated randomly, and thus may have different global maxima, making it difficult to properly compare the organizational performance across different simulation runs (e.g. the performance of 0.79 in a landscape with global maximum of 0.80 might be more desirable than the performance of 0.89 in a landscape with global maximum of 1.0). To ensure the proper comparison we normalize each performance value by the global maximum of the landscape.} and synchrony for $T=500$ periods. We look at the long-run and short-run values for these two metrics, defining them as the mean values over $T=500$ and $T=100$ periods respectively. Thus, we obtain 4 vectors with $R=1000$ observations, and use them to perform our hypothesis tests.

\section{Results}
\label{sec:results}

In this section we present the results of our simulations. We start with the baseline scenarios to check the validity of our model compared to the existing research. We then look at the main findings regarding the role of conformity in addressing the trade-off between performance and synchrony. Finally, we discuss the robustness of our simulation model.

\subsection{The trade-off between performance and synchrony}
\label{sec:baseline}

First, we want to verify that the results generated by our simulations are aligned with the findings established in the literature. Particularly, we expect the decentralization to have a positive overall effect on the organizational performance (the effect can be crowded out by the presence of interdependencies between the units) \citep{siggelkow05a,harrington00} and we expect to observe the trade-off between performance and synchrony (see Sec. \ref{sec:tradeoff}), given our specific formulation of the synchrony based on counting bits. To do that, we run simulations for scenarios with varying degrees of centralization, ignoring the conformity motive of the unit managers. For each such scenario we calculate the mean organizational performance and synchrony over an observation span of $T=500$ periods and compare these values. Finally, we check if the findings still hold in presence of interdependencies between the units.

\begin{figure}[!tb]
	\includegraphics[width=\linewidth]{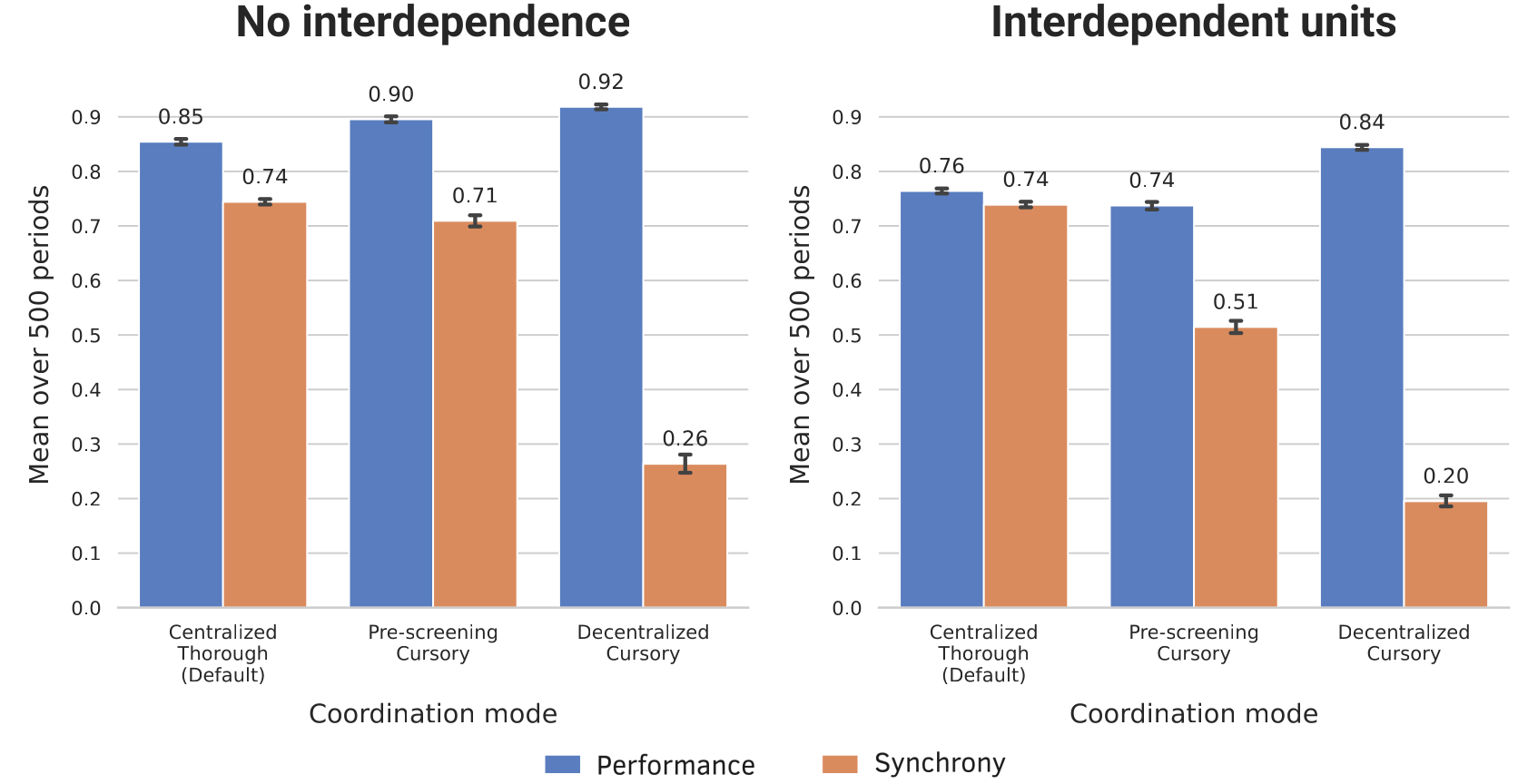}
	\caption{Performance and Synchrony trade-off in the absence of conformity}
	\label{fig:baseline_barplot}
	\subnote{The error bars represent the $99\%$ confidence interval}
\end{figure}

We can see in Figure \ref{fig:baseline_barplot} that the default organizational performance (0.85) increases with the decentralization, as the branch managers are given the opportunity to pre-screen the alternatives (0.90) or even further --- to make fully autonomous decisions (0.92). This effect is less pronounced when the branches are interdependent, as expected. At the same time, we can see how the synchrony drastically falls with the decentralization (e.g. from 0.74 to 0.26), mainly due to the fact that in this formulation the unit managers are only interested in their own unit's performance, while the upper management has synchrony among its objectives. It can be argued that the units facing highly correlated task environments should arrive at similar decisions even without any coordination. However, in our model the agents face the \textit{rugged} landscapes (see Sec. \ref{sec:taskenv}) that can have multiple high performing solutions (local optima) for the same sets of tasks, which might result in low synchrony, unless the upper management specifically aims for high synchrony.

%
%

Table \ref{tab:effectsize} presents these results in more detail, by taking the centralized coordination mode with thorough search as the default, and performing pairwise comparisons with the other coordination modes. Specifically, it compares the mean of performance and synchrony in the short-run ($T=100$ periods) and the long-run ($T=500$ periods), using the Mann-Whitney U-test \citep{mannwhitney1947} to check whether they are significantly different and the Cliff's $\delta$ \citep{cliff1993} to measure the effect size. The table illustrates the trade-off between performance and synchrony using the negative effect sizes (marked in red) that accompany the largest effect sizes of each column (marked in bold), which also happen to correspond to the decentralized coordination mode (e.g. $(+0.74, \redd{-0.90})$), thus emphasizing the positive effect of decentralization on performance. Although, the coordination mode with pre-screening by the unit managers and a thorough search by the upper management represents a middle ground by slightly but significantly increasing both the performance and synchrony (e.g. $(+0.42,+0.14)$), it fails to do so if the units are interdependent (e.g. $(\redd{-0.20}, \redd{-0.81})$).

Up to this point, we have been able to establish the overall positive role of decentralization on performance, to operationalize the concept of organizational synchrony in the context of the $N\!K$-landscapes and confirm that it exhibits the tradeoff with performance.

\begin{table*}[!tb]
	\small
	\singlespacing
	\caption{Mann-Whitney U-test and Cliff's $\delta$ effect size for coordination modes compared to the default (no conformity)}
	\label{tab:effectsize}
	\begin{tblr}{
			colspec={lX[c]X[c]X[c]X[c]X[c]X[c]X[c]X[c]},
			cell{1}{1}={r=3}{c},
			cell{1}{2,6}={c=4}{c},
			cell{2}{2,4,6,8}={c=2}{c},
			cell{even[4-14]}{1}={r=2}{l}
		}
		\toprule
		{\bf Coordination\\mode}  & \bf No interdependence & & & & \bf Interdependent units & & & \\
		\cmidrule[lr]{2-5} \cmidrule[lr]{6-9}
		& Long-run & & Short-run & & Long-run & & Short-run & \\
		\cmidrule[lr]{2-3} \cmidrule[lr]{4-5} \cmidrule[lr]{6-7} \cmidrule[lr]{8-9}
		& Perf. & Sync. & Perf. & Sync. & Perf. & Sync. & Perf. & Sync. \\
		\midrule
		{Centralized\\Thorough\\(Default)} & \lgry{$0.0$} & \lgry{$0.0$} & \lgry{$0.0$} & \lgry{$0.0$}  & \lgry{$0.0$} & \lgry{$0.0$} & \lgry{$0.0$} & \lgry{$0.0$} \\
		& \lgry{-} & \lgry{-} & \lgry{-} & \lgry{-} & \lgry{-} & \lgry{-} & \lgry{-} & \lgry{-} \\
		{Centralized\\Cursory}    & \redd{$-0.16^{***}$} & \redd{$-0.28^{***}$} & \redd{$-0.13^{***}$} & \redd{$-0.34^{***}$} & \redd{$-0.18^{***}$} & \redd{$-0.20^{***}$} & \redd{$-0.10^{***}$} & \redd{$-0.32^{***}$} \\
		& $(0.03)$      & $(0.02)$      & $(0.03)$      & $(0.02)$      & $(0.03)$      & $(0.03)$      & $(0.03)$      & $(0.02)$      \\
		{Pre-screening\\Thorough} & $+0.42^{***}$ & $+0.14^{***}$ & $+0.68^{***}$ & $+0.44^{***}$ & \redd{$-0.20^{***}$} & \redd{$-0.81^{***}$} & $+0.11^{***}$ & \redd{$-0.64^{***}$} \\
		& $(0.02)$      & $(0.03)$      & $(0.02)$      & $(0.02)$      & $(0.03)$      & $(0.02)$      & $(0.03)$      & $(0.02)$      \\
		{Pre-screening\\Cursory}  & $+0.40^{***}$ & \lgry{$+0.04$} & $+0.63^{***}$ & $+0.30^{***}$ & \redd{$-0.15^{***}$} & \redd{$-0.82^{***}$} & $+0.13^{***}$ & \redd{$-0.68^{***}$} \\
		& $(0.02)$      & \lgry{$(0.03)$}& $(0.02)$      & $(0.03)$      & $(0.03)$      & $(0.01)$      & $(0.03)$      & $(0.02)$      \\
		{Decentralized\\Thorough} & $\mbf{+0.74^{***}}$ & \redd{$-0.90^{***}$} & $\mbf{+0.91^{***}}$ & \redd{$-0.80^{***}$} & $+0.51^{***}$ & \redd{$-0.98^{***}$} & $\mbf{+0.56^{***}}$ & \redd{$-0.98^{***}$} \\
		& $(0.02)$      & $(0.01)$      & $(0.01)$      & $(0.02)$      & $(0.02)$      & $(0.00)$      & $(0.02)$      & $(0.00)$      \\
		{Decentralized\\Cursory}  & $+0.63^{***}$ & \redd{$\mbf{-0.93^{***}}$} & $+0.85^{***}$ & \redd{$\mbf{-0.86^{***}}$} & $\mbf{+0.70^{***}}$ & \redd{$\mbf{-0.99^{***}}$} & $+0.53^{***}$ & \redd{$\mbf{-0.99^{***}}$} \\
		& $(0.02)$      & $(0.01)$      & $(0.01)$      & $(0.01)$      & $(0.02)$      & $(0.00)$      & $(0.02)$      & $(0.00)$      \\
		\bottomrule
	\end{tblr}

	\subnote{$^{*} p < 0.05$; $^{**} p < 0.01$; $^{***} p < 0.001$ according to the Mann-Whitney U-test that the performance or synchrony of a coordinaion mode of interest is not significantly different from that of the default (Centralized Thorough). The values show the Cliff's $\delta$ effect size (and its standard error in parentheses), which can take values from $-1$ to $+1$, where the values closer to these boundaries refer to the probability of any random observation from a sample of interest being greater (or less, if the sign is negative) than any random observation from the sample of the default coordination mode, while the values closer to $0$ refer to no statistically significant difference between the samples of interest. The greatest effect sizes for each envirionment are highlighted in bold (and in red for negative effect sizes) and the cases with large p-values are de-emphasized using light gray color.}
\end{table*}

\subsection{Enabling decentralization with conformity}
\label{sec:mainfind}

%
%

\begin{figure}[!tb]
	\includegraphics[width=\linewidth]{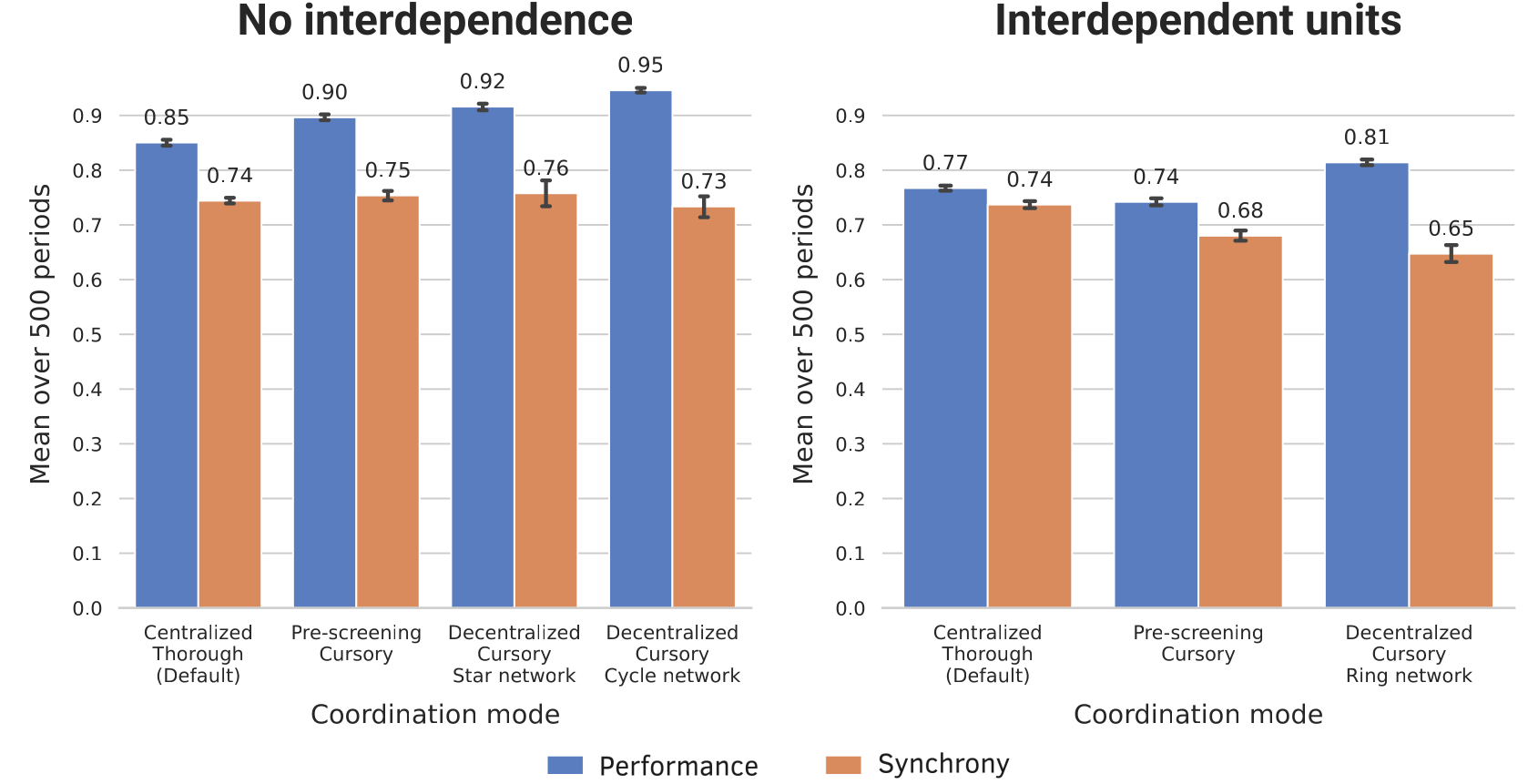}
	\caption{Performance and Synchrony trade-off in presence of conformity}
	\label{fig:conformity_barplot}
	\subnote{The error bars represent the $99\%$ confidence interval}
\end{figure}

Having validated our baseline model against the established findings, we now want to see how the results might change if we incorporate into it the unit managers' willingness to conform. Particularly, we are interested if conformity might resolve the trade-off between performance and synchrony. We perform a very similar analysis to the one in the previous section, but this time we also look at the different network structures through which the unit managers communicate.

Figure \ref{fig:conformity_barplot} shows the mean organizational performance and synchrony over $T=500$ periods for centralized, pre-screening, and (the network structures with highest synchrony of) decentralized coordination mode. We can see that the positive effect of decentralization on the performance (which is less pronounced when the units are interdependent) still exists when the unit managers are willing to conform to each other (e.g. 0.85 to 0.95 for Cycle network). However, the synchrony (0.74) no longer falls drastically, as we give more autonomy to the unit managers. In fact, it slightly increases when the unit managers are given the ability to pre-screen the alternatives (0.75), or to make fully aunomous decisions (e.g. 0.76 for Star network when all unit managers conform to one central unit). This shows that the unit managers' decentralized willingness to adjust their behavior to match that of others, leads to synchrony even without upper management's centralized decision to pursue synchrony as an objective. When the units are interdependent, however, the trade-off persists, and the decentralization decreases the synchrony (e.g. from 0.74 to 0.65 for Ring network), albeit to a smaller extent than without conformity (0.20).

\begin{table*}[!tb]
	\small
	\singlespacing
	\caption{Mann-Whitney U-test and Cliff's $\delta$ effect size for coordination modes compared to the default (conformity)}
	\label{tab:effectsizeconf}
	\begin{tblr}{
	   		colspec={lX[c]X[c]X[c]X[c]X[c]X[c]X[c]X[c]},
	   		cell{1}{1}={r=3}{c},
	   		cell{1}{2,6}={c=4}{c},
	   		cell{2}{2,4,6,8}={c=2}{c},
	   		cell{even[4-26]}{1}={r=2}{l}
	   		}
	   	\toprule
		{\bf Coordination\\mode}  & \bf No interdependence & & & & \bf Interdependent units & & & \\
		\cmidrule[lr]{2-5} \cmidrule[lr]{6-9}
		  & Long-run & & Short-run & & Long-run & & Short-run & \\
	   	\cmidrule[lr]{2-3} \cmidrule[lr]{4-5} \cmidrule[lr]{6-7} \cmidrule[lr]{8-9}
	 	  & Perf. & Sync. & Perf. & Sync. & Perf. & Sync. & Perf. & Sync. \\
	   	\midrule
		{Centralized Thorough\\ Line network (Default)} & \lgry{$0.0$} & \lgry{$0.0$} & \lgry{$0.0$} & \lgry{$0.0$}  & \lgry{$0.0$} & \lgry{$0.0$} & \lgry{$0.0$} & \lgry{$0.0$} \\
				    			  			  	& \lgry{-} & \lgry{-} & \lgry{-} & \lgry{-} & \lgry{-} & \lgry{-} & \lgry{-} & \lgry{-} \\
		{Centralized Thorough\\Other networks} 	& \lgry{$-0.04^{}$} & \lgry{$+0.02^{}$} & \lgry{$-0.01^{}$} & \lgry{$-0.01^{}$} & \lgry{$+0.01^{}$} & \lgry{$-0.00^{}$} & \lgry{$-0.01^{}$} & \lgry{$-0.00^{}$} \\
											   	& \lgry{$(0.02)$} & \lgry{$(0.02)$} & \lgry{$(0.02)$} & \lgry{$(0.02)$} & \lgry{$(0.02)$} & \lgry{$(0.02)$} & \lgry{$(0.02)$} & \lgry{$(0.02)$} \\
		{Centralized Cursory\\All networks} 	& \redd{$-0.18^{***}$} & \redd{$-0.26^{***}$} & \redd{$-0.10^{***}$} & \redd{$-0.34^{***}$} & \redd{$-0.15^{***}$} & \redd{$-0.20^{***}$} & \redd{$-0.10^{***}$} & \redd{$-0.31^{***}$} \\
												& $(0.02)$ & $(0.02)$ & $(0.02)$ & $(0.02)$ & $(0.02)$ & $(0.02)$ & $(0.02)$ & $(0.02)$ \\
		{Pre-screening Cursory\\Line network}	& $+0.36^{***}$ & $+0.25^{***}$ & $+0.61^{***}$ & $+0.46^{***}$ & \redd{$-0.18^{***}$} & \redd{$-0.42^{***}$} & \lgry{$+0.04^{}$} & \redd{$-0.21^{***}$} \\
												& $(0.02)$ & $(0.03)$ & $(0.02)$ & $(0.02)$ & $(0.03)$ & $(0.02)$ & \lgry{$(0.03)$} & $(0.03)$ \\
		{Pre-screening Cursory\\Cycle network}  & $+0.39^{***}$ & $+0.30^{***}$ & $+0.61^{***}$ & $+0.53^{***}$ & \redd{$-0.17^{***}$} & \redd{$-0.26^{***}$} & \lgry{$+0.03^{}$} & \redd{$-0.07^{**}$} \\
												& $(0.02)$ & $(0.03)$ & $(0.02)$ & $(0.02)$ & $(0.03)$ & $(0.03)$ & \lgry{$(0.03)$} & $(0.03)$ \\
		{Pre-screening Cursory\\Ring network} 	& $+0.40^{***}$ & $\mbf{+0.41^{***}}$ & $+0.65^{***}$ & $\mbf{+0.65^{***}}$ & \redd{$-0.18^{***}$} & \lgry{$-0.04^{}$} & \lgry{$+0.04^{}$} & $+0.24^{***}$ \\
												& $(0.02)$ & $(0.02)$ & $(0.02)$ & $(0.02)$ & $(0.03)$ & \lgry{$(0.03)$} & \lgry{$(0.03)$} & $(0.02)$ \\
		{Pre-screening Cursory\\Star network}	& $+0.41^{***}$ & $+0.35^{***}$ & $+0.64^{***}$ & $+0.61^{***}$ & \redd{$-0.15^{***}$} & \redd{$-0.14^{***}$} & $+0.06^{*}$ & $+0.13^{***}$ \\
												& $(0.02)$ & $(0.02)$ & $(0.02)$ & $(0.02)$ & $(0.03)$ & $(0.03)$ & $(0.03)$ & $(0.03)$ \\
		{Decentralized Cursory\\Line network} 	& $+0.61^{***}$ & \lgry{$-0.04^{}$} & $+0.85^{***}$ & \redd{$-0.27^{***}$} & $+0.40^{***}$ & \redd{$\mbf{-0.85^{***}}$} & $+0.53^{***}$ & \redd{$\mbf{-0.90^{***}}$} \\
												& $(0.02)$ & \lgry{$(0.03)$} & $(0.01)$ & $(0.03)$ & $(0.02)$ & $(0.02)$ & $(0.02)$ & $(0.01)$ \\
		{Decentralized Cursory\\Cycle network}  & $\mbf{+0.77^{***}}$ & $+0.08^{**}$ & $\mbf{+0.89^{***}}$ & \redd{$-0.25^{***}$} & $+0.33^{***}$ & \redd{$-0.63^{***}$} & $+0.50^{***}$ & \redd{$-0.81^{***}$} \\
												& $(0.01)$ & $(0.03)$ & $(0.01)$ & $(0.03)$ & $(0.02)$ & $(0.02)$ & $(0.02)$ & $(0.02)$ \\
		{Decentralized Cursory\\Ring network}	& $+0.61^{***}$ & \redd{$-0.11^{***}$} & $+0.82^{***}$ & \redd{$-0.07^{**}$} & $\mbf{+0.43^{***}}$ & \redd{$-0.31^{***}$} & $\mbf{+0.57^{***}}$ & \redd{$-0.49^{***}$} \\
												& $(0.02)$ & $(0.03)$ & $(0.01)$ & $(0.03)$ & $(0.02)$ & $(0.03)$ & $(0.02)$ & $(0.02)$ \\
		{Decentralized Cursory\\Star network}   & $+0.51^{***}$ & $+0.10^{***}$ & $+0.75^{***}$ & $+0.22^{***}$ & $+0.33^{***}$ & \redd{$-0.64^{***}$} & $+0.48^{***}$ & \redd{$-0.72^{***}$} \\
												& $(0.02)$ & $(0.03)$ & $(0.02)$ & $(0.03)$ & $(0.02)$ & $(0.02)$ & $(0.02)$ & $(0.02)$ \\
		\bottomrule
	\end{tblr}

	\subnote{$^{*} p < 0.05$; $^{**} p < 0.01$; $^{***} p < 0.001$ according to the Mann-Whitney U-test that the performance or synchrony of a coordinaion mode of interest is not significantly different from that of the default (Centralized Thorough). The values show the Cliff's $\delta$ effect size (and its standard error in parentheses), which can take values from $-1$ to $+1$, where the values closer to these boundaries refer to the probability of any random observation from a sample of interest being greater (or less, if the sign is negative) than any random observation from the sample of the default coordination mode, while the values closer to $0$ refer to no statistically significant difference between the samples of interest. The greatest effect sizes for each envirionment are highlighted in bold (and in red for negative effect sizes) and the cases with large p-values are de-emphasized using light gray color. Thourough search cases for Pre-screening and Decentralized coordination mode have been omitted due to having no significant difference with the Cursory search (see Appendix \ref{sec:appendixB} for more details).}
\end{table*}

In Table \ref{tab:effectsizeconf} we can see that in centralized coordination mode, changing the network structures has no significant effect on performance and synchrony (the second row is marked in light gray, which denotes high p-value for the Mann-Whitney U-test), primarily because the upper management's centralized decision is unaffected by the unit managers communication patterns. We can also see that the coordination mode in which the unit managers are allowed to pre-screen the alternatives, increases both the performance and synchrony in the long-run (e.g. (+0.41, +0.35) for Star network), and to an even greater extent in the short-run (e.g. (+0.64, +0.61) for Star network). For the decentralized coordination mode, the Cycle network leads to the highest increase in performance both in the long-run ($+0.77$) and in the short-run ($+0.89$) at the expense of short-term decline in synchrony (\redd{$-0.25$}), while the Star network completely resolves the trade-off by increasing performance and synchrony both in the short-run and the long-run (see the last row).

When the units are interdependent, the pre-screening coordination mode decreases the long-run performance and synchrony (e.g. $(\redd{-0.17}, \redd{-0.26})$ for Cycle network), while not significantly affecting the short-run performance. The short-run synchrony, though, increases for Ring ($+0.24$) and Star ($+0.13$) networks. Similarly, for the decentralized coordination the increase in performance is accompanied with the (sometimes drastic) decline in synchrony (e.g. \redd{$-0.85$} in the long-run and \redd{$-0.90$} in the short-run for Line network), with the Ring network leading to the smallest decline (\redd{$-0.31$}).

In this section we have shown that conformity might help to resolve the trade-off between performance and synchrony for some network structures by enabling decentralization, but fails to do so when the units are interdependent.

\subsection{Organizations with more units}

Thus far, we have presented the results of simulations of organizations with $P=5$ units. As explained in Sec. \ref{sec:process}, we aimed for an accurate comparison between different simulation runs, and to achieve that we normalized the (organizational and unit) performances by the maximum possible performance in that run. To do that, we computed global maximum of every (randomly) generated performance landscape, which is computationally expensive, and for this reason we have chosen $P=5$ units as the default. However, for the sake of completeness, we also simulated organizations with a greater number of units, dropping the requirement to normalize the performances. By not calculating the global maxima of landscapes we were able to significantly reduce the computational costs of our simulations. This simplification, by design, decreases the accuracy of the analysis of performance (although the large number of simulation runs is expected to offset the loss in accuracy), but does not negatively affect the analysis of synchrony, since it was already not being normalized.

\begin{figure}[!tb]
	\includegraphics[width=\linewidth]{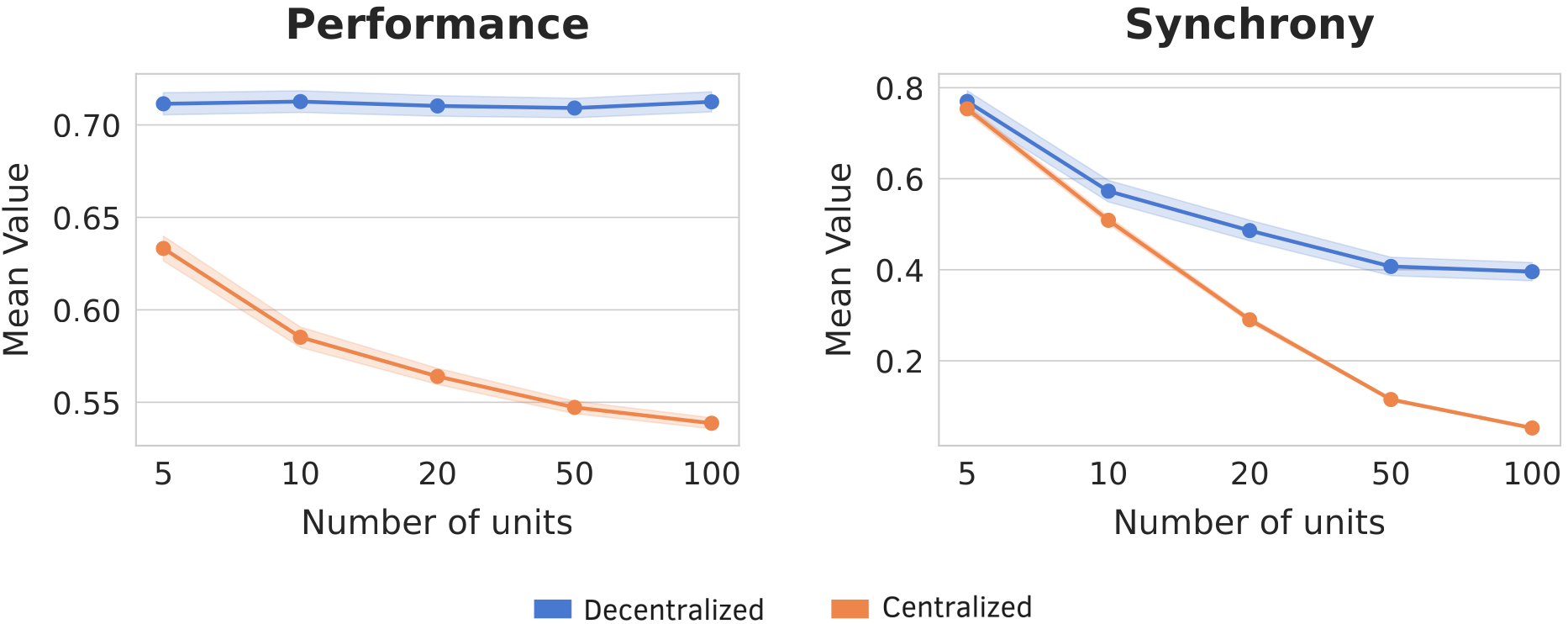}
	\caption{Comparing coordination modes for different numbers of units}
	\label{fig:ofat_p}
	\subnote{One-factor-at-a-time sensitivity analysis in $P$ with $99\%$ confidence intervals, the number of units in the organization. For computational feasibility, this analysis has been performed without normalizing the performances by the global maximum in their corresponding landscapes, which explains why the values for $P=5$ might not match the findings reported in the previous sections. This limitation does not apply to the values of synchrony.}
\end{figure}

We performed a \textit{one-factor-at-a-time (OFAT)} analysis by comparing the mean long-run performance and synchrony of a centralized organization with thorough search and a decentralized organization with Cycle network structure in presence of conformity. Figure \ref{fig:ofat_p} shows how our results are even more pronounced for organizations with a greater number of units. In particular, as the number of units increases, the organizational performance falls for centralized coordination mode but is unaffected for the decentralizated coordination mode. The organizational synchrony declines for both coordination modes as the number of units increases, although, the decline is significantly stronger for the centralized mode. As a result, the decentralization increases both the performance and synchrony, thus resolving the trade-off even for larger organizations.

%

\subsection{Robustness checks}

Our main findings claim that in presence of conformity, the decentralized coordination mode leads to higher performance and synchrony compared to the centralized coordination mode. However, our model defines scenarios using many parameters, and, while we have carefully picked them in our analysis, we want to check whether our findings are robust to the changes in these parameters. To achieve that, we perform (local) OFAT analysis, similarly to the previous section, and the (global) sensitivity analysis using the Sobol' method, as suggested by \citet{ten2016}.

In OFAT analysis, we identify a default set of parameters (see Online Appendix 2) and vary one parameter at a time, holding others fixed. For each variation we compare performance and synchrony values of the decentralized and centralized coordination modes, and check if their ordinal relationship still holds. In Online Appendix 2 we present the results, which show that decentralization leads to higher performance than centralization for different variations in the correlation between landscapes ($\rho$), memory span of unit managers ($T_M$), upper management's goals for performance and synchrony ($g_\Phi, g_\mathbf{S}$), and unit managers' weights for performance and conformity ($w_\phi, w_\kappa$), except for the extreme case when the unit managers do not care about performance at all ($w_\phi=0$). For synchrony, the findings are robust for agents' memory span $T_M$ shorter than 50 periods and all values of upper management's goal for synchrony $g_\mathbf{S}$. The findings do not hold if the correlation between landscapes $\rho$ is less than $0.9$, the upper management's goal for performance $g_\Phi$ is less than $1$, and the unit managers' weight for conformity $w_\kappa$ is less than $0.5$.

Finally, we perform a global sensitivity analysis using Sobol' method to check which parameters explain the most variance in the performance and synchrony (see Online Appendix 3). We find that the highest first-order effects are the weight $w_\kappa$ unit managers' put into conformity, and the coordination mode, whereas the interdependence between units has low first-order effects. Moreover, the goal for performance has the first-order effect on performance but has no effect on synchrony, and vice versa. Most of the effects, however, are the total effects, which means that the parameters have less direct effect on our model than an indirect effect through the interdependencies between other parameters, which is plausible due to the design of the model.

\section{Discussion}
\label{sec:discussion}

The previous section reported the results of simulations that we ran to address the question of whether it is possible to use the individuals' willingness to conform to their peers to achieve organizational performance and synchrony without a centralized directive (which is a default in multi-unit organizations). To directly answer this question, we abstracted away some details such as costs of promoting norms for sharing and the emergence process of conformity. Thus, taking conformity as given, we compared the organizational performance and synchrony for three coordination modes with varying levels of centralization and four network structures with different levels of connectedness and the direction of information flow.

\subsection{Effects of conformity on performance and synchrony}

We found that when the unit managers conform to each other, decentralizing the decision-making can indeed increase both performance and synchrony if they are communicating over the Star network structure and, in the long-run, also over the Cycle network structure. When the units are interdependent, though, the Centralized (default) coordination mode remains the best way to achieve high synchrony, whereas the decentralization leads to a higher performance at the cost of decline in synchrony (with the Ring network leading to a smaller decline and higher performance than other network structures). 

Cycle network leads to a higher performance than Star network because it features a decentralized flow of information leading to a more diverse set of decisions being shared, increasing the likelihood of containing the higher performing ones, akin to \textit{parallel search} \citep{kavadias2009,baumann18}. On the other hand, Cycle network leads to lower synchrony than Star network and even a decline in synchrony in the short-run because it does not feature direct connectedness, which slows down the flow of information needed to achieve synchrony \citep{khodzhimatov2023}. Practically, a Decentralized multi-unit organization with conforming unit managers and Star network is very similar to the Centralized organization in which the upper management orders the units to adopt the solutions that increase performance and synchrony for the entire organization. Interestingly, the unit managers' willingness to increase the performance of their own unit and to conform to one of their peers, leads to both higher organizational performance and higher synchrony than the upper management's direct mandate to do so. With this finding we expanded the scope of problems that decentralization solves better than the centralization \citep{harrington00,siggelkow05a} to conditionally include the pursuit of synchrony.

When the units are interdependent, which corresponds to the notion of \textit{turbulent environments} \citep{siggelkow05a}, it becomes harder to find the high performing solution that is also replicable, because the mere act of replicating it, might reduce its performance. This, in turn, leads to lower synchrony, because the unit managers will less likely settle for a fixed solution that could then be replicated. Ring network in this scenario leads to the smallest decline in synchrony because it features the diversity of shared solutions (due to its decentralized nature) and the fast information flow (due to its high connectedness), both of which are important when the environment is turbulent and the solutions become obsolete very fast.

\subsection{Effects of coordination mode on performance and synchrony}

For completeness of our analysis, we deemed it necessary to clarify some of the results that appear unintuitive. First of all, we saw that the decentralized coordination mode improves performance over the (default) centralized mode even when the units are interdependent, while the established research shows that the former will fail to capture the externalities and lead to lower performance \citep{kauffman91,ethiraj04}. This is explained by the fact that the upper management in our model always pursues two goals (goal of performance and goal of synchrony) simultaneously, which leads to a lower performance, in favor of the higher synchrony.

Another observation that needs to be addressed is why the pre-screening coordination mode, which constitutes the middle ground between full centralization and decentralization, does not result in the middle values between the two. Particularly, we need to explain why the pre-screening mode leads to higher synchrony than both centralized and decentralized modes when the units are not interdependent. The reason lies in the design of our model, where the upper management only checks several random combinations from the alternatives that may or may not be pre-screened by the unit managers. Thus, when the pre-screening takes place, the upper management chooses among the alternatives that already increase the unit managers' conformity, which increases the chance of picking the one that increases the organizational synchrony and lowers the chance of encountering alternatives that neither increase performance nor synchrony and thus would get ignored. Similarly, when the units are interdependent, the pre-screening results in the unit managers submitting the alternatives that would increase their unit's performance, which decreases the chance of encountering the combinations that increase organizational performance while sacrificing the performance of some individual units.

\subsection{Our research in context}

Multi-unit organizations, which constitute an interesting combination of large multidivisional organizations and a set of independent businesses operating in the same market, have been subject to multiple empirical and qualitative studies \citep{jensen2007, seran2023}. \citet{harrington00} were among the first to approach this topic using the simulation, and found that when the units are facing similar external conditions, they can benefit from the collective search for the better performing solutions, with the highest performing solution found by one of the units being observed by the central management and enforced onto all the other units. They noticed that this similarity in practices (which we called \textit{synchrony} in this paper) is beneficial for the organization, but they could not offer another mechanism apart from the centralized enforcement of this similarity. They did propose some version of team-based incentive schemes or giving vetoing power to the upper management, but these were the formal mechanisms that did not take the individuals' behavioral and psychological considerations in mind, which was one of the criticisms raised by \citet{gavetti2007}, when they were talking about the future directions the organizational science must take. 

Our research tried to address this gap and include the insights from psychology and behavioral sciences regarding the willingness of individuals to conform to their peers \citep{cialdini04,ajzen91} into the discussion of the multi-unit organizations. We found that, if organizations manage to promote the norms for sharing knowledge and construct the appropriate network structures, they can reach high performance and synchrony without the need for centralized control.

\section{Conclusion}
\label{sec:conclusion}

In this paper we explored the centralization vs. decentralization dilemma that multi-unit organizations face, where the former allows them to reach higher synchrony between units and the latter allows the units to adapt to their local environments and increase the performance. We have proposed that, given appropriately designed communication channels and the culture and norms that promote sharing, the individuals' inner willingness to conform to their peers can itself increase the organizational synchrony without the centralized order. We have built an agent-based model of a multi-unit organization with different coordination modes and network structures, and performed simulations that confirmed our proposed findings.

One of the limitations of our study is that we modeled the individual utility as the weighted sum of performance and conformity, whereas we realized later on that using goal programming would be more appropriate for modeling the decision making of individuals. This did not affect the results of this paper, because we used the equal weights in our additive utility function, which technically corresponds to the goals of $1$ given to performance and conformity, but it was reflected in the sensitivity analysis. Additionally, we abstracted away the costs of promoting the norms for sharing knowledge and of establishing the network structures, and the noise and error that can occur in communication.

Overall, though, when building the model we used the best practices and tools found in the literature. We modeled the coordination modes using the methodology of \citet{siggelkow05a}. We abandoned the usual notion of using one performance landscape in which multiple agents search for solutions and imitate others, in favor of correlated landscapes which can also interact with each other using the methodology of \citet{kauffman91} and \citet{verel13}. We increased the number of agents and the number of decisions that each agent faces in other papers \citep{harrington00,siggelkow05a}, which allowed us to operationalize the concept of synchrony by counting the similar bits.

As a result, we came up with a working model (the codes are well-documented and are available in the online materials) which can be extended to explore multiple new directions, such as an in-depth network analysis of the synchrony in organizations, the addition of the complex contagions aspect to adopting the knowledge, and testing exciting new ways to coordinate decision making, such as letting one unit explore for several periods and then share their knowledge with others.


\bibliographystyle{elsarticle-harv} 
\bibliography{refs.bib}

\appendix

\begin{appendix}
	
\section{Model description}
\label{sec:appendixA}

\begin{figure}[!htb]
	\includegraphics[width=0.8\linewidth]{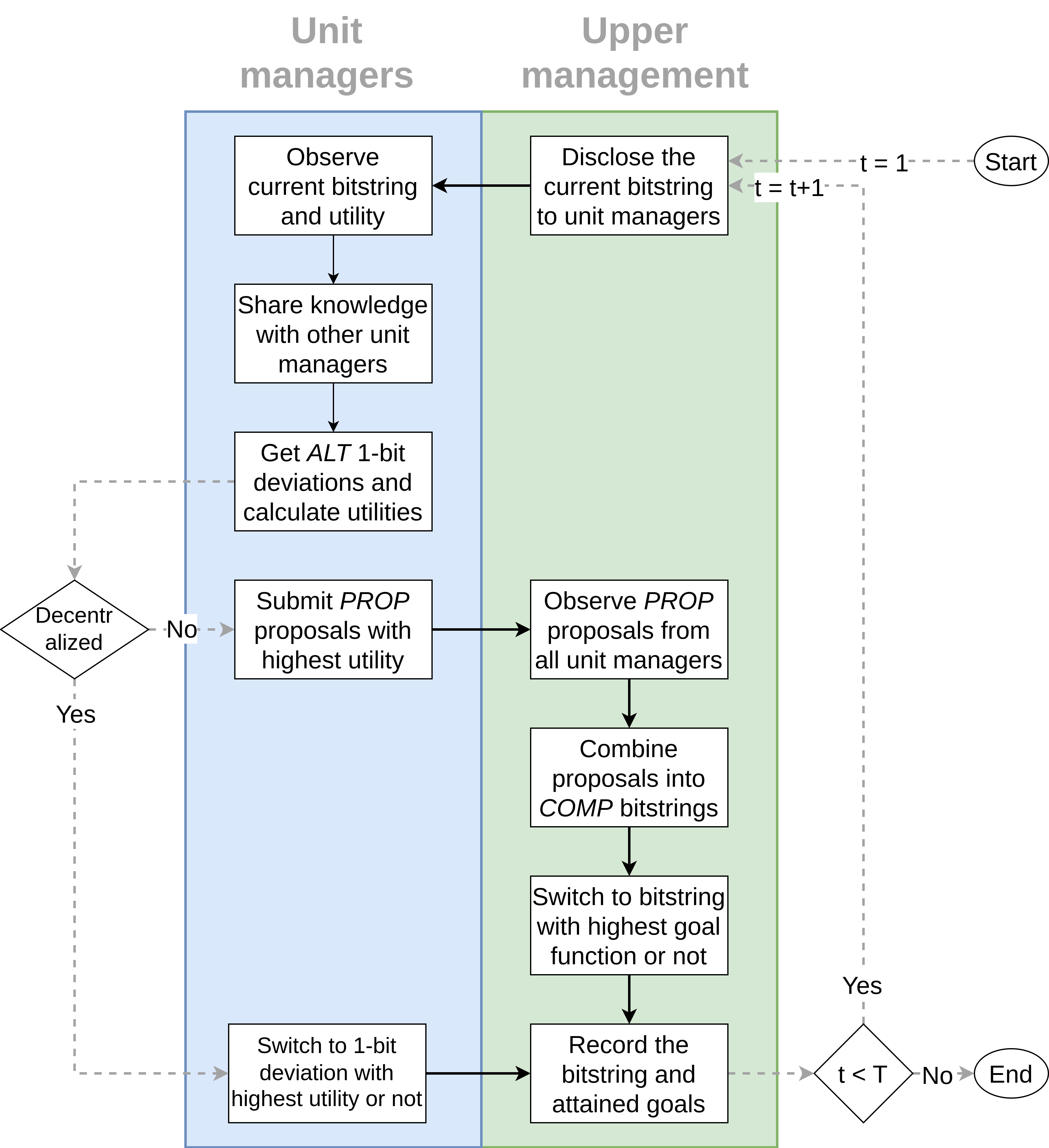}
	\caption{Process overview and scheduling}
	\label{fig:process}
\end{figure}

\clearpage
\section{Effects of thorough search}
\label{sec:appendixB}

\begin{table}[!hp]
	\small
	\caption{Mann-Whitney U-test and Cliff's $\delta$ effect size for coordination modes compared to the default (conformity)} \label{tab:effectsizeconfthor}
		\begin{tblr}{
			colspec={lX[c]X[c]X[c]X[c]X[c]X[c]X[c]X[c]},
			cell{1}{1}={r=3}{c},
			cell{1}{2,6}={c=4}{c},
			cell{2}{2,4,6,8}={c=2}{c},
			cell{even[4-26]}{1}={r=2}{l}
		}
		\toprule
		{\bf Coordination\\mode}  & \bf No interdependence & & & & \bf Interdependent units & & & \\
		\cmidrule[lr]{2-5} \cmidrule[lr]{6-9}
		& Long-run & & Short-run & & Long-run & & Short-run & \\
		\cmidrule[lr]{2-3} \cmidrule[lr]{4-5} \cmidrule[lr]{6-7} \cmidrule[lr]{8-9}
		& Perf. & Sync. & Perf. & Sync. & Perf. & Sync. & Perf. & Sync. \\
		\midrule
		{Centralized Thorough\\ Line network (Default)} & \lgry{$0.0$} & \lgry{$0.0$} & \lgry{$0.0$} & \lgry{$0.0$}  & \lgry{$0.0$} & \lgry{$0.0$} & \lgry{$0.0$} & \lgry{$0.0$} \\
		& \lgry{-} & \lgry{-} & \lgry{-} & \lgry{-} & \lgry{-} & \lgry{-} & \lgry{-} & \lgry{-} \\
		{Pre-screening Thorough\\Line network}	& $+0.44^{***}$ & $+0.35^{***}$ & $+0.70^{***}$ & $+0.60^{***}$ & \redd{$-0.16^{***}$} & \redd{$-0.38^{***}$} & $+0.11^{***}$ & \redd{$-0.11^{***}$} \\
												& $(0.02)$ & $(0.03)$ & $(0.02)$ & $(0.02)$ & $(0.03)$ & $(0.02)$ & $(0.03)$ & $(0.03)$ \\
		{Pre-screening Thorough\\Cycle network} & $+0.43^{***}$ & $+0.37^{***}$ & $+0.69^{***}$ & $+0.63^{***}$ & \redd{$-0.17^{***}$} & \redd{$-0.13^{***}$} & $+0.07^{**}$ & $+0.14^{***}$ \\
												& $(0.02)$ & $(0.02)$ & $(0.02)$ & $(0.02)$ & $(0.03)$ & $(0.03)$ & $(0.03)$ & $(0.03)$ \\
		{Pre-screening Thorough\\Ring network} 	& $+0.43^{***}$ & $\mbf{+0.43^{***}}$ & $+0.69^{***}$ & $+0.70^{***}$ & \redd{$-0.15^{***}$} & \lgry{$+0.01^{}$} & $+0.10^{***}$ & $+0.36^{***}$ \\
												& $(0.02)$ & $(0.02)$ & $(0.02)$ & $(0.02)$ & $(0.03)$ & \lgry{$(0.03)$} & $(0.03)$ & $(0.02)$ \\
		{Pre-screening Thorough\\Star network} 	& $+0.38^{***}$ & $\mbf{+0.43^{***}}$ & $+0.66^{***}$ & $\mbf{+0.72^{***}}$ & \redd{$-0.14^{***}$} & \redd{$-0.11^{***}$} & $+0.10^{***}$ & $+0.22^{***}$ \\
												& $(0.02)$ & $(0.02)$ & $(0.02)$ & $(0.02)$ & $(0.03)$ & $(0.03)$ & $(0.03)$ & $(0.03)$ \\

		{Decentralized Thorough\\Line network} 	& $+0.69^{***}$ & \lgry{$-0.01^{}$} & $+0.90^{***}$ & \redd{$-0.15^{***}$} & $+0.36^{***}$ & \redd{$\mbf{-0.78^{***}}$} & $+0.52^{***}$ & \redd{$\mbf{-0.85^{***}}$} \\
												& $(0.02)$ & \lgry{$(0.03)$} & $(0.01)$ & $(0.03)$ & $(0.02)$ & $(0.02)$ & $(0.02)$ & $(0.01)$ \\
		{Decentralized Thorough\\Cycle network} & $\mbf{+0.85^{***}}$ & $+0.19^{***}$ & $\mbf{+0.95^{***}}$ & \lgry{$-0.02^{}$} & $+0.26^{***}$ & \redd{$-0.47^{***}$} & $+0.48^{***}$ & \redd{$-0.65^{***}$} \\
												& $(0.01)$ & $(0.03)$ & $(0.01)$ & \lgry{$(0.03)$} & $(0.02)$ & $(0.03)$ & $(0.02)$ & $(0.02)$ \\
		{Decentralized Thorough\\Ring network}	& $+0.77^{***}$ & \lgry{$-0.03^{}$} & $+0.92^{***}$ & \lgry{$+0.03^{}$} & $\mbf{+0.41^{***}}$ & \redd{$-0.16^{***}$} & $\mbf{+0.57^{***}}$ & \redd{$-0.21^{***}$} \\
												& $(0.01)$ & \lgry{$(0.03)$} & $(0.01)$ & \lgry{$(0.03)$} & $(0.02)$ & $(0.03)$ & $(0.02)$ & $(0.03)$ \\
		{Decentralized Thorough\\Star network}	& $+0.61^{***}$ & $+0.07^{**}$ & $+0.83^{***}$ & $+0.21^{***}$ & $+0.29^{***}$ & \redd{$-0.45^{***}$} & $+0.47^{***}$ & \redd{$-0.53^{***}$} \\
												& $(0.02)$ & $(0.03)$ & $(0.01)$ & $(0.03)$ & $(0.02)$ & $(0.03)$ & $(0.02)$ & $(0.02)$ \\
		\bottomrule
	\end{tblr}
	
	{\textit{Notes:} $^{*} p < 0.05$; $^{**} p < 0.01$; $^{***} p < 0.001$ according to the Mann-Whitney U-test that the performance or synchrony of a coordinaion mode of interest is not significantly different from that of the default (Centralized Thorough). The values show the Cliff's $\delta$ effect size (and its standard error in parentheses), which can take values from $-1$ to $+1$, where the values closer to these boundaries refer to the probability of any random observation from a sample of interest being greater (or less, if the sign is negative) than any random observation from the sample of the default coordination mode, while the values closer to $0$ refer to no statistically significant difference between the samples of interest. The greatest effect sizes for each envirionment are highlighted in bold (and in red for negative effect sizes) and the cases with large p-values are de-emphasized using light gray color.}
\end{table}

\end{appendix}

\end{document}